\documentclass[twocolumn,amsmath,amssymb,showpacs,preprintnumbers]{revtex4}

\usepackage{graphicx}
\usepackage{bm}
\usepackage{multirow}

\begin{document}

\title{Influence of magnetic-field inhomogeneity on nonlinear magneto-optical resonances}

\author{S.~Pustelny}
\affiliation{Centrum Bada\'n Magnetooptycznych, M.~Smoluchowski
Institute of Physics, Jagiellonian University, Reymonta 4, 30-059
Krak\'ow, Poland}
\author{D.~F.~Jackson~Kimball}
\affiliation{Department of Physics, California State University --
East Bay, 25800 Carlos Bee Blvd., Hayward, CA 94542, USA}
\author{S.~M.~Rochester}
\affiliation{Department of Physics, University of California at
Berkeley, Berkeley, CA 94720-7300, USA}
\author{V.~V.~Yashchuk}
\affiliation{Advanced Light Source Division, Lawrence Berkeley
National Laboratory, Berkeley CA 94720, USA}
\author{D.~Budker}
\affiliation{Department of Physics, University of California at
Berkeley, Berkeley, CA 94720-7300, USA} \affiliation{Nuclear Science
Division, Lawrence Berkeley National Laboratory, Berkeley CA 94720,
USA}

\begin{abstract}

    In this work, a sensitivity of the rate of relaxation of
    ground-state atomic coherences to magnetic-field inhomogeneities is
    studied. Such coherences give rise to many interesting phenomena in
    light-atom interactions, and their lifetimes are a limiting factor
    for achieving better sensitivity, resolution or contrast in many
    applications. For atoms contained in a vapor cell, some of the
    coherence-relaxation mechanisms are related to magnetic-field
    inhomogeneities. We present a simple model describing relaxation due
    to such inhomogeneities in a buffer-gas-free anti-relaxation-coated
    cell. A relation is given between relaxation rate and magnetic-field
    inhomogeneities including the dependence on cell size and atomic
    spices. Experimental results, which confirm predictions of the
    model, are presented. Different regimes, in which the relaxation
    rate is equally sensitive to the gradients in any direction and in
    which it is insensitive to gradients transverse to the bias magnetic
    field, are predicted and demonstrated experimentally.

\end{abstract}

 \maketitle

\section{Introduction}

In recent years, there has been a considerable interest in many
physical phenomena associated with the existence of coherence
between atomic states. Such coherences, induced and detected by
light, form the basis of certain nonlinear optical effects and are
essential in such applications as magnetometry
\cite{Alexandrov2004,Giles2001,Weis1998,Budker1998,Budker2000,FMNMOR},
electromagnetically induced transparency
\cite{Arimondo1996,Harris1997}, and quantum gates
\cite{Turchette1995}. They are also extensively employed in tests of
fundamental symmetries (see, for example, reviews
\cite{AlexandrovReview,RevModPhys}) and in frequency standards
\cite{Vanier2005}.

The lifetime of atomic coherences involving excited atomic states is
generally limited to twice the time required for spontaneous
emission of a photon and transition to a lower state. On the other
hand, for coherences between atomic ground states the effective
coherence lifetime $\tau$ is either determined by the interaction
time between the light and atoms, or by the time between
coherence-destroying collisions. In many applications, the longer
the lifetime of the coherences, the better resolution, contrast or
sensitivity that can be achieved.

In a typical experiment, involving a single light beam and a glass
cell containing only alkali-metal vapor at low pressure, the
effective coherence lifetime is given by the transit time of the
atoms through the light beam. In order to suppress
coherences-destroying collisions of atoms with cell walls and
increase the lifetime $\tau$, one of two methods is employed. The
first method is to add to the cell a buffer (usually noble) gas at
relatively high pressure. Since in the first approximation the
collisions with the buffer gas are elastic, $\tau$ is then given by
the time for the alkali atoms to diffuse from the light beam. The
second method is to apply an anti-relaxation coating to the inner
walls of the cell, preventing spin-depolarizing collisions of the
atoms with the walls. This allows atoms to leave the light beam and
later return to it with the coherences intact. Using these methods
the coherence lifetimes have been prolonged to hundreds of
milliseconds (see, for example Refs.
\cite{BudkerAntirelaxation,Erhard2001} and references therein). As
wall relaxation is decreased, however, other sources of relaxation
become important, such as spin-exchange self-collisions and
magnetic-field gradients, of which the latter is discussed here.

The sensitivity of the rate of relaxation $\gamma$ of the
ground-state coherences ($\gamma=1/\tau$) to magnetic-field
inhomogeneities was previously studied under different experimental
conditions \cite{Watanabe1977,Cates1988,McGregor1990,Bohler1994}. In
a series of papers \cite{Watanabe1977} the sensitivity of the
relaxation rate to the magnetic-field gradients was studied in
anti-relaxation-coated cells. Theoretical predictions supported by
numerical simulations were compared with data obtained in a
high-resolution Zeeman spectroscopy experiment performed with
relatively strong bias magnetic field $B=50$ G. In such range of
magnetic fields, the nonlinear Zeeman effect significantly
contributes to relaxation and it cannot be neglected. In Refs.\
\cite{Cates1988,McGregor1990,Bohler1994} the sensitivity of the
relaxation rate $\gamma$ to the magnetic-field gradients in a
buffer-gas cell was studied both theoretically and experimentally.

In this work the rate of relaxation due to magnetic-field
inhomogeneity is studied in buffer-gas-free anti-relaxation-coated
cells at relatively low magnetic fields ($|B|<160$ mG), where the
influence of the nonlinear Zeeman effect on relaxation is
negligible. A naive theoretical model of the sensitivity of the
relaxation rate to the magnetic-field inhomogeneity is given. A
relation between the sensitivity of the relaxation rate and cell
size or atomic spice contained in the cell is derived.
Experimentally, the problem is studied using nonlinear
magneto-optical rotation with frequency-modulated light (FM NMOR)
\cite{FMNMOR}. The sensitivity of the FM NMOR resonances to
first-order magnetic-field gradients is analyzed and compared with
the model predictions. The sensitivity to the magnetic-field
gradients is studied for different bias magnetic fields. This
enables observation of two different regimes, one in which the rate
of relaxation $\gamma$ depends equally on the magnetic-field
inhomogeneities in each direction and the other, in which it is
completely insensitive to the transverse inhomogeneities.

The article is organized as follows. In Section \ref{sec:theory} the
theoretical model of the sensitivity of the rate of relaxation to
the magnetic-field inhomogeneities is given. The experimental
apparatus and the measurement technique are described in Section
\ref{sec:setup}. In Section \ref{sec:results} the experimental
results are presented and compared with the predictions of the
model. Conclusions are summarized in Section \ref{sec:conclusion}.

\section{Theoretical model} \label{sec:theory}

Consider atoms contained in a buffer-gas-free,
anti-relaxation-coated spherical cell of radius $R$. Since at room
temperature at saturated alkali-vapor densities an atom's mean free
path is on the order of hundreds of meters, the atoms travel freely
between collisions with the cell walls. To analyze the influence of
magnetic-field inhomogeneities on the rate of relaxation of the
ground-state coherences, we use a simple model for a first-order
magnetic-field gradient, for example, $\partial B_i/\partial x_i$,
in which the gradient field in each half of the cell is replaced by
a constant magnetic field $\Delta B_i$ in that half. In other words,
the cell is considered to have the magnetic field $\Delta
B_i=(3R/8)(\partial B_i/\partial x_i)$ for $x_i>0$ and the opposite
field $-\Delta B_i$ for $x_i<0$. At this point we assume that
$\Delta B_i$ is the only magnetic field inside the cell. However,
the effect of homogenous bias magnetic field
$\boldsymbol{\mathrm{B}}$ is considered below.

Between wall collisions, the atomic spins precess due to the
magnetic field and acquire an average phase
\begin{equation}
    \phi_{grad}\approx\frac{g\mu_B\Delta B_i\tau_c}{\hbar},
    \label{eq:phasegradient}
\end{equation}
where $g$ is the Land\'{e} factor, $\mu_B$ is the Bohr magneton, and
$\tau_c\approx4R/3v$ is the average time between two collisions and
$v$ is the r.m.s.\ atomic thermal velocity. However, since acquiring
the phase is a random process the actual phases acquired by the atom
vary between collisions.

Another source of random phase is inelastic wall collisions. A
simple model of wall-collision relaxation is the following. During a
collision with the wall, an atom is stuck to the surface for
approximately $10^{-10}$ s \cite{Bouchiat1966} (note that the
duration of an elastic collision is $\sim$10$^{-12}$ s), and is then
released in a random direction with a random velocity. During the
time it spends on the surface the atom feels an excess magnetic
field $B_{wall}$. This field causes the atomic spin to rotate,
producing an average phase shift $\phi_{wall}$. Combining the phases
due to the magnetic-field inhomogeneity and wall collisions for $N$
successive bounces in quadrature, the total phase is
\begin{equation}
    \left(\phi_{total}\right)^2
    =N\left[\left(\phi_{grad}\right)^2+\left(\phi_{wall}\right)^2\right].
    \label{eq:phasesinglepass}
\end{equation}

A characteristic relaxation time corresponds to a decrease of
initial spin polarization by a factor of $e$. One can show that this
happens when $\phi_{total}=\sqrt{2}$, and thus the total number of
wall collisions before dephasing $N_\gamma$ is
\begin{equation}
    N_\gamma=
    \frac{\sqrt{2}}{\left(\phi_{grad}\right)^2+\left(\phi_{wall}\right)^2}.
    \label{eq:totalphase}
\end{equation}
The relaxation rate $\gamma$ of the atoms can be written as
\begin{equation}
    \gamma=\frac{1}{N_\gamma \tau_c}.
    \label{eq:relaxationconstantbase}
\end{equation}
In the two limiting cases in which only one mechanism of relaxation
is present, Eq.\ (\ref{eq:relaxationconstantbase}) takes the forms
\begin{eqnarray}
    \gamma_{grad}=\frac{1}{N_{grad}\tau_c},
    \label{eq:relaxationconstantgradient}\\
    \gamma_{wall}=\frac{1}{N_{wall}\tau_c},
    \label{eq:relaxationconstantwall}
\end{eqnarray}
where $N_{grad}$ and $N_{wall}$ are respectively the numbers of
bounces before relaxation when only the gradient relaxation or the
wall relaxation is present. It is noteworthy that including other
relaxation mechanisms such as spin-exchange collisions does not
change the present treatment.

Combining Eqs.\ \eqref{eq:phasegradient}, (\ref{eq:totalphase}),
(\ref{eq:relaxationconstantbase}),
(\ref{eq:relaxationconstantgradient}), and
(\ref{eq:relaxationconstantwall}) we find
\begin{equation}
\begin{split}
    \gamma
    &{}\approx\frac{4\mu_B^2g^2R\Delta B_i^2}{3\sqrt{2}\hbar^2v}+\gamma_{wall}\\
    &{}=\xi g^2R^3\left(\frac{\partial B_i}{\partial x_i}\right)^2+\gamma_{wall},
    \label{eq:relaxationconstantdeltaB}
\end{split}
\end{equation}
where $\xi=3\mu_B^2/(16\sqrt{2}\hbar^2v)$. One sees that the rate of
relaxation due to magnetic-field gradients depends quadratically on
the inhomogeneity of the magnetic field $(\partial B_i/\partial
x_i)^2$, scales as $R^3$, and depends on the atomic species through
the $g^2$ dependence.

Now, we consider the effect of the homogenous bias magnetic field
$\boldsymbol{\mathrm{B}}$, leading to a Larmor precession frequency
$\Omega_L=g\mu_BB$. In this situation we consider two regimes: (1)
$\Omega_L R/v\ll 1$, in which atomic spins rotate by only a small
angle between collisions with the cell walls and (2) $\Omega_L
R/v\gg 1$, in which the spins rotate by a large angle between
successive wall bounces. In the first regime, since small rotations
commute, i.e., the result of the composite rotation does not depend
on the order of rotations, the spins' precession around orthogonal
components of the magnetic field can be, to a good approximation,
considered as independent. Thus, comparable sensitivity to
longitudinal and transverse magnetic-field gradients can be
expected. In the second regime the Larmor precession is rapid and
the strong bias magnetic field breaks the symmetry of the system.
Since fields transverse to the strong bias fields are only second
order corrections to the total magnetic field, the sensitivities of
atomic-polarization relaxation rates to longitudinal and transverse
magnetic-field gradients are different.

In the limit of high bias field, one would expect the relaxation
rate to become completely insensitive to transverse gradients.
However, according to Maxwell's equations generation of a
magnetic-field gradient in one direction requires gradients in other
directions ($\nabla\cdot\boldsymbol{\mathrm{B}}=0$). Thus, in
conditions of our experiment, when a gradient transverse to the
light propagation direction is applied (say $\partial B_y/\partial
y$), a gradient $\partial B_x/\partial x$ given by
\begin{equation}
    \frac{\partial B_x}{\partial x}=
    -\frac{1}{2}\frac{\partial B_y}{\partial y}.
    \label{eq:transversegradientproduction}
\end{equation}
also appears along the longitudinal direction $x$. Since the
relaxation is quadratic in the magnetic-field gradient, it is
expected that the relaxation rate should be four times as sensitive
to gradients nominally along the longitudinal direction as they are
to those nominally in a transverse direction.

\section{Experimental setup} \label{sec:setup}

\begin{figure}
    \includegraphics{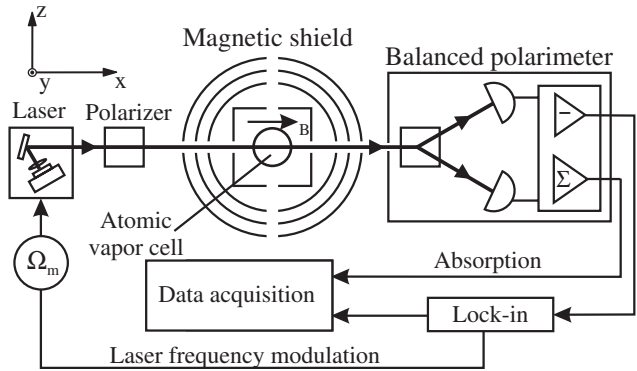}
    \caption{Experimental setup. The magnetic-field coils enabling
    generation of bias magnetic field along $x$ and compensation of
    the residual fields in other directions, as well as gradient
    coils, are not shown.} \label{fig:setup}
\end{figure}

Relaxation rates may be studied by observing the widths of
resonances in optical rotation using the FM NMOR technique
\cite{FMNMOR}. The layout of the experimental setup is shown in
Fig.\ \ref{fig:setup}.
Rubidium atoms are contained in anti-relaxation-coated
buffer-gas-free spherical vapor cells of different diameters and
containing different isotopic compositions of rubidium (Table
\ref{tab:cells}) \footnote{The cells used in this work were
previously intensively studied in different experiments. A detailed
analysis of hyperfine and Zeeman relaxations in these cells is
presented in Ref. \cite{BudkerAntirelaxation}.}.
\begin{table}[h]
\caption{\label{tab:cells}Anti-relaxation-coated, buffer-gas-free
vapor cells used in the present work.}
\begin{ruledtabular}
\begin{tabular}{c|c|c}
    Cell designation & Outer diameter (cm) & Isotope \\
    \hline \hline Ale10 & 10.0(1) & $^{85}$Rb \\
    \hline Rb10 & 10.2(1) & $^{87}$Rb \\
    \hline Gibb & 10.3(3) & Natural Rb \\
    \hline H2 & 3.4(1) & $^{87}$Rb \\
\end{tabular}
\end{ruledtabular}
\end{table}
A cell is placed inside a four-layer magnetic shield providing
passive attenuation of the DC magnetic fields to the level of one
part per 10$^6$ \cite{YashchukShield}. A set of three mutually
orthogonal magnetic-field coils is mounted inside the innermost
shielding layer. These coils are used for compensation of the
residual magnetic field inside the shield, as well as for generating
a bias magnetic field along the $x$ axis. Data were taken for the
bias magnetic field ranging from 0.2 mG to 155 mG. An additional set
of three calibrated coils is used for compensation and generation of
first-order magnetic-field gradients inside the shield.

The rubidium atoms interact with $y$-polarized light produced by an
external-cavity diode laser operating at the rubidium $D1$ line (795
nm). The  $\sim$3 $\mu$W light beam is 2 mm in diameter and
propagates along $x$. The laser light is frequency modulated at a
rate $\Omega_m\approx 2\Omega_L$ with a modulation depth of 300 MHz
(peak to peak). The central frequency of the laser is tuned to the
low-frequency wing of the $F=2\rightarrow F'=1$ transition for
$^{87}$Rb measurements and to the center of the $F=3\rightarrow F'$
transition group for $^{85}$Rb measurements, in order to produce the
maximum FM NMOR signal in each case. The central frequency of the
laser is stabilized with a dichroic atomic vapor lock
\cite{DAVLLWieman,BudkerDAVLL} modified for operation with
frequency-modulated light. The rotation of the polarization plane of
the light transmitted through the vapor cell is analyzed with a
balanced polarimeter (a crystalline polarizer rotated by 45$^\circ$
in the $yz$-plane and two photodiode detectors). A photodiode
difference signal is detected with a lock-in amplifier at the first
harmonic of $\Omega_m$. In-phase and quadrature components of the
detected signal are stored with a computer.

Some of the results presented in this paper were recorded with an
experimental arrangement slightly different from Fig.
\ref{fig:setup}, essentially the same as the one described in Ref.\
\cite{Pustelny2006}. Instead of one light beam, an additional
unmodulated (probe) laser is used. The probe laser operating at the
rubidium $D2$ line (780 nm) is tuned to the center of the
$F=2\rightarrow F'$ transition group of $^{87}$Rb. It is polarized
in the $y$ direction, that is, in the same direction as the pump
beam. The probe laser light propagates along $x$, while the
frequency-modulated (pump) laser beam propagates along $z$. The
polarization-plane rotation of the probe laser beam was measured
with the polarimeter at the first harmonic of the pump-laser
modulation frequency. Despite the changes in the experimental
arrangement the experimental results obtained in the one- and
two-beam experiments are consistent.

\section{Results and Analysis} \label{sec:results}

\begin{figure}
    \includegraphics{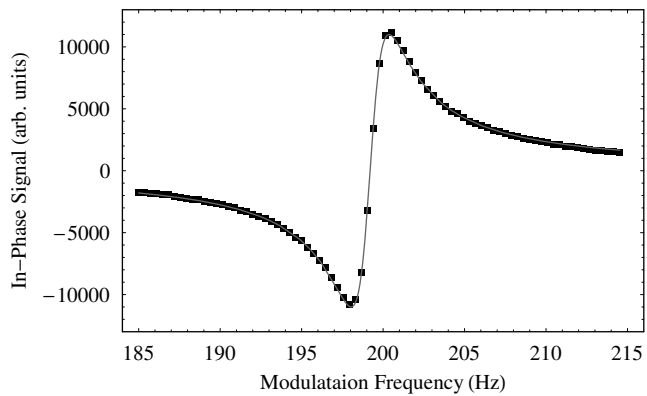}
    \caption{A typical in-phase FM NMOR signal recorded in
    Rb10 cell. Square points represents experimental data.
    The solid line is a dispersive Lorentzian fit. The data
    were recorded in a single-beam experiment with the light power 4 $\mu$W
    and magnetic field corresponding to $\Omega_L\approx 2\pi\cdot 100$ Hz.} \label{fig:FMNMORsignal}
\end{figure}

A typical in-phase FM NMOR signal recorded as a function of a
modulation frequency is shown in Fig.\ \ref{fig:FMNMORsignal}.
Experimental data were fit with a dispersive Lorenzian. The width of
the FM NMOR signal $\Delta\nu$, which corresponds to the relaxation
rate of the ground-state coherences $\gamma$ by the relation
$\gamma=2\pi\cdot\Delta\nu$, is half of the distance between two
peaks in the signal.

\begin{figure}
    \includegraphics{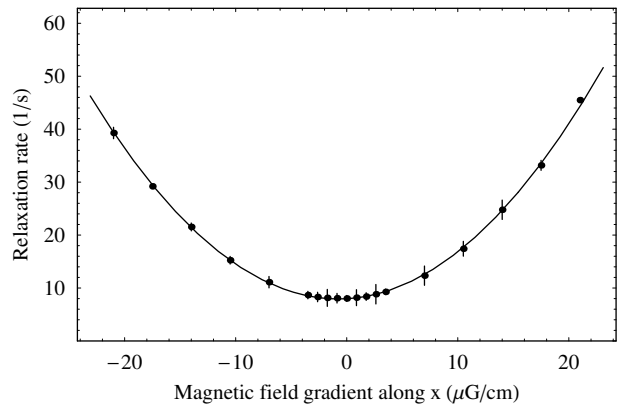}
    \caption{The relaxation rate of the ground-state coherences,
    extracted from the width of the FM NMOR resonance, vs. the first-order
    magnetic-field gradient applied using $x$-oriented coils. Note
    that the application of such a gradient is accompanied by
    appearance of gradients in $y$ and $z$ directions (see text).
    The experimental results are fit with a quadratic dependence
    [Eq. (\ref{eq:fitting})]. The signals were recorded in a single-beam
    arrangement in the Rb10 cell with bias magnetic field
    $B\approx250$ $\mu$G and light power 4 $\mu$W.}
    \label{fig:dependence}
\end{figure}

In Fig.\ \ref{fig:dependence} the dependence of the rate of
relaxation of the ground-state coherences $\gamma$ is presented as a
function of the magnetic-field gradients applied with $x$ coils.
In order to verify the predictions of the model the experimental
data were fit with the quadratic dependence
\begin{equation}
    \gamma=
    a_i\left(\frac{\partial B_i}{\partial x_i}\right)^2
    +\gamma_0,
    \label{eq:fitting}
\end{equation}
where $a_i$ is the coefficient describing the sensitivity to the
magnetic-field gradient applied using $x_i$-oriented coils and
$\gamma_0$ is the relaxation rate in the absence of the gradients.
As seen in Fig.\ \ref{fig:dependence} the experimental data are in
good agreement with the theoretically predicted quadratic dependence
of the relaxation rate as a function of the magnetic-field gradient.
The agreement was also observed for two transverse directions $y$
and $z$.

\begin{figure}
    \includegraphics{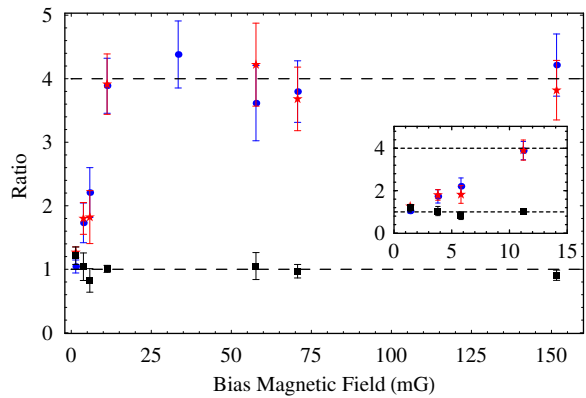}
    \caption{(Color online) Ratio between sensitivity to the
    magnetic-field gradients generated with $x$, $y$ and $z$
    coils. Circles correspond to ratio between sensitivities
    applied with $x$ and $y$ coils ($a_x/a_y$), stars with $x$ and $z$
    coils ($a_x/a_z$), and squares with $y$ and $z$
    coils ($a_y/a_z$). The data were recorded in two-beam arrangement in H2
    cell. The pump- and probe-beam powers were 15 $\mu$W and 5 $\mu$W,
    respectively.} \label{fig:biasfielddependence}
\end{figure}

The sensitivities to the magnetic-field gradients were also studied
as a function of the strength of the bias magnetic field. The ratios
between sensitivities to gradients applied with $x$, $y$, and $z$
coils are shown in Fig.\ \ref{fig:biasfielddependence}.
The sensitivity to the gradients applied with either of the
transverse coils is the same ($a_y/a_z\approx 1$) over the whole
range of bias magnetic fields, as expected by symmetry. However, as
predicted in Sec.\ \ref{sec:theory}, the ratio between the
sensitivity to magnetic-field gradients applied with the
longitudinal and transverse coils changes with the strength of the
bias magnetic field. In the zero-field limit of the bias field the
sensitivity to the magnetic-field gradients applied with the
longitudinal and either of the transverse coils is the same
($a_x/a_y\rightarrow 1$ and $a_x/a_z\rightarrow 1$). For stronger
bias fields the ratio between the sensitivity to the first-order
magnetic-field gradients applied with the longitudinal and
transverse coils increases until it levels off at $a_x/a_y\approx
a_x/a_z\approx 4$ for $B>15$ mG. These results are in agreement with
the theory, which predicts $a_\text{long}/a_\text{trans}=1$ for
$\Omega_L R/v\ll 1$, and $a_\text{long}/a_\text{trans}=4$ for
$\Omega_L R/v\gg 1$. For the experimental conditions of Fig.\
\ref{fig:biasfielddependence}, 15 mG corresponds to
$\Omega_LR/v\approx 3$. Thus the experimental results confirm that
for high bias fields the rate of relaxation $\gamma$ is insensitive
to the transverse part of the magnetic-field inhomogeneities.

Table \ref{tab:sensitivity} gives the sensitivity of the rate of
relaxation of the ground-state coherences to magnetic-field
gradients applied with each of the three gradient coils for the four
cells studied here.
\begin{table}[t]
    \caption{The sensitivity of the relaxation rate to the magnetic
    field gradients applied in a given direction. The bias magnetic
    field was $B\approx 250$ $\mu$G, except for Ale10, for which it was
    $B\approx 300$ $\mu$G. In Gibb cell, the sensitivity to the
    magnetic-field gradients was measured for $^{87}$Rb.}
    \begin{ruledtabular}
    \begin{tabular}{c|c|c|c}
        Cell & $a_x$ (cm$^2$/s$\,\mu$G$^2$) & $a_y$ (cm$^2$/s$\,\mu$G$^2$) & $a_z$
        (cm$^2$/s$\,\mu$G$^2$)\\
        \hline \hline Rb10 & 83.6(7) & 50.9(7) & 47.7(7) \\
        Gibb & 89.2(13) & 59.7(25) & 52.8(13) \\
        Ale10 & 32.0(7) & 18.8(13) & 18.2(7) \\
        H2 & 2.2(4) & 1.4(2) & 1.6(2) \\
    \end{tabular}
    \end{ruledtabular}
    \label{tab:sensitivity}
\end{table}
The sensitivity to the magnetic-field gradients varies with the
orientation of the coils used for generation of the gradients, cell
size and rubidium isotope.

As seen in Table \ref{tab:sensitivity}, the sensitivity to the
gradients in larger cells is stronger than in the smaller cell.
According to Eq.\ (\ref{eq:relaxationconstantdeltaB}), the
sensitivity to the magnetic-field gradients scales as $R^3$. In
order to check this, Table \ref{tab:sensitivityratios} gives the
experimentally measured ratios for cells of different sizes along
with the theoretical predictions. To calculate the theoretical ratio
between sensitivity due to different sizes of the cells their inner
radii were used. They were estimated by subtraction of a wall
thickness, which were assumed to be $\sim 0.2$ cm, from the cells'
outer radii.
\begin{table}[h]
    \caption{\label{tab:sensitivityratios}Ratio between the
    sensitivities to the magnetic-field gradients applied with coils
    oriented in a given direction for cells of different radii (upper
    subtable) and cells containing different isotopes of rubidium (lower subtable).
    The uncertainties in the ratio between sensitivity to the gradients
    in cells of different sizes represent deviations
    from a perfect spherical shape and uncertainty in estimation of
    cell thickness. To calculate the ratios for different isotopes the
    slight differences in the cell internal sizes were taken into account.}
    \begin{ruledtabular}
    \begin{tabular}{c|c|c|c|c}
        Cells & $x$ axis & $y$ axis & $z$ axis & Theory\\
        \hline \hline \multicolumn{5}{c}{Different radii} \\ \hline \hline
        Rb10/H2 & 40(8) & 43(8) & 33(4) & 35(8)\\
        Gibb/H2 & 38(8) & 36(6) & 30(5) & 36(10)\\
        Rb10/Gibb & 0.94(3) & 0.85(5) & 0.90(4) & 0.94(12)\\
        \hline \hline \multicolumn{5}{c}{Different isotopes} \\ \hline \hline
        Ale10/Gibb & 0.39(2) & 0.34(4) & 0.38(3) & \multirow{2}*{0.44}\\
        Ale10/Rb10 & 0.41(2) & 0.39(4) & 0.41(3) \\
    \end{tabular}
    \end{ruledtabular}
\end{table}
The experimental data are consistent with the predictions of the
model.

According to the model, the FM NMOR width scales with Land\'e factor
as $g^2$ [Eq.\ (\ref{eq:relaxationconstantdeltaB})]. For the two
isotopes of rubidium, $^{85}$Rb and $^{87}$Rb, for which the Land\'e
factors are 1/3 and 1/2, respectively, the expected ratio is 4/9.
Rough agreement is seen between the experimental results and
predictions for the ratios of sensitivities for cells containing
different isotopes [Table \ref{tab:sensitivityratios}]. The results
in Table \ref{tab:sensitivityratios} are scaled to take into account
the different cell sizes. For the results relating the Ale10 and
Gibb cells, we associate a difference from the theoretical value
with the slightly nonspherical shape of the Gibb cell (in addition
to the overall non-sphericity, it does not have a typical stem but
it has a number of tubulations). Another source of deviation is the
different bias field used for the measurements with the Ale10 cell,
as noted in the caption to Table \ref{tab:sensitivity}. As discussed
above, this difference affects the sensitivities to the transverse
magnetic-field gradients.

\section{Conclusion} \label{sec:conclusion}

We have presented a simple model describing a relation between the
relaxation rate of the ground-state coherences $\gamma$ of atoms
contained in buffer-gas-free anti-relaxation-coated cell and
magnetic-field inhomogeneities. The results of the experiments using
nonlinear magneto-optical rotation with frequency-modulated light
have confirmed the model across the board. We showed that the rate
of relaxation $\gamma$ of the ground-state Zeeman coherences is
proportional to the square of the magnetic-field inhomogeneity
(first-order magnetic-field gradients), and that it scales as the
cube of the cell size and as the square of the Land\'e factor.
Additionally, we provide experimental evidence that the sensitivity
to the longitudinal part of the magnetic-field inhomogeneity is
independent of bias magnetic field, but the sensitivity to the
transverse part of the inhomogeneity changes with bias field. At
small bias fields the sensitivity to transverse inhomogeneities is
similar to the sensitivity in the longitudinal direction, but at
larger fields (where the Zeeman frequency exceeds the wall-collision
rate), it vanishes.

\begin{acknowledgments}
The authors would like to acknowledge H. Robinson for providing one
of the cells and E. B. Alexandrov, W. Gawlik, J. Higbie, M.
Ledbetter, M. V. Romalis, and I. Savukov for helpful discussions.
This work is supported by DOD MURI grant \# N-00014-05-1-0406, KBN
grant \# 1 P03B 102 30, and a NSF US-Poland collaboration grant. One
of the author (S.P.) is a scholar of the project co-financed from
the European Social Fund.
\end{acknowledgments}

\end{document}